\documentstyle[12pt,prc,aps,epsf]{revtex}

\begin{document}


\title{Attractive Central Potential in the SU(3) Skyrme Model}
\author{Hideyuki Abe
\thanks{Present address : Kachida-minami 2-8-23, Tsuzuki-ku, Yokohama 224, Japan.}
}
\address{Institute of Physics, University of Tokyo, \\
Komaba, Tokyo 153, Japan}

\maketitle

\begin{abstract}
The interaction between the hyperon and the nucleon is investigated in the SU(3) Skyrme model.
The static potential,
which is expanded in terms of the modified SU(3) rotation matrices,
is obtained for several orientations with the Atiyah-Manton ansatz.
The interaction is calculated for the NN, $\Lambda$N, and $\Sigma$N systems.
The medium-range attraction of the central potential between $\Lambda$ and N
is obtained by considering the $\Lambda$-$\Sigma$ mixing
through the intermediate state.
\end{abstract}

\pacs{12.39.Dc, 13.75.Cs, 13.75.Ev}

\section{Introduction}

It is widely accepted
that quantum chromodynamics (QCD) is the fundamental theory
for the strong interaction.
The high-energy behavior of QCD is well described by the perturbed QCD
because of its asymptotic freedom.
It explains the experimental data
such as those in the deep inelastic scattering process.
On the other hand,
it is difficult to describe the low energy properties of QCD
because the effective coupling constant
increases with the decreasing momentum
as the renormalization group analysis suggests.

'tHooft proposed to use the inverse number of colors $1/N_{\rm C}$ as an expansion parameter
by generalizing QCD to the SU($N_{\rm C}$) gauge theory\cite{tHft}.
In the large $N_{\rm C}$ limit, it becomes the theory of the weakly interacting meson.
Witten pointed out that the baryons should appear as topological solitons\cite{Wit}.
The Skyrme model is recognized
as an effective theory of QCD in the large $ N_{\rm C} $ limit
although the Skyrme lagrangian is not derived from QCD directly\cite{Sky}.
The baryon number is introduced into the Skyrme model from a topological point of view.
Skyrme proposed a stable configuration called the hedgehog ansatz.
The quantization is performed by introducing the collective coordinates,
the flavor rotation of the hedgehog configuration\cite{ANW}.
The Skyrme model explains the static properties of the baryon
such as the charge radius and the magnetic moment
with 30\% accuracy.

The product ansatz is a two-skyrmion configuration proposed by Skyrme\cite{Sky}.
It is a good approximation
so far the two skyrmions are separated in the long distance.
The numerical simulation is a direct method
to obtain the exact two-baryon configuration\cite{VWWW}.
There is another way to describe the skyrmion configuration
with a few parameters.
The Atiyah-Manton ansatz is constructed
from the instanton configuration in the SU(2) gauge theory\cite{AM,HOA}.
The stable configuration with the torus shape
can be described by this ansatz.
Even if the skyrmion configuration is obtained,
there is a problem that the attraction in the central potential
at the intermediate range is absent.
It is being solved by considering
the N-$\Delta$ mixing through the intermediate state\cite{Saito,DePace},
the finite-$N_{\rm C}$ effect\cite{JJP},
the higher-order terms generated by $ \omega$-meson,
and the radial excitation.
By diagonalizing the potential between the NN and N$\Delta$ states for each channel,
one can construct the better eigenstate.
It amounts to the N-$\Delta$ mixing.
The finite $N_{\rm C}$ effects are often considered together with the NN-N$\Delta$ mixing.
Since the Skyrme model is recognized as an effective theory in the large $ N_{\rm C} $ limit,
the finite $N_{\rm C}$ correction is required\cite{WWWA}.

The Skyrme model is extended into the SU(3) flavor symmetry.
There are two approaches to deal with the extra strange degrees of freedom.
One is the bound state approach\cite{CK}
in which the symmetry breaking is regarded as large.
The K-meson is introduced as a small fluctuation
from the SU(2) symmetry.
Another is the collective coordinate method
which is based on the SU(3) symmetry.
In this method, the symmetry breaking is taken to be small.
The symmetry breaking is treated perturbatively\cite{MNP,Chem}.
Yabu and Ando unified these two approaches by the exact treatment
of the symmetry breaking\cite{YA}.
Yabu-Ando approach reproduces the mass splitting
of the baryons in the same multiplet.
In the two-baryon case, only the product ansatz has been investigated
because of the complexity of the numerical simulation of the SU(3) Skyrme model\cite{KE,SSG}.

In this paper, we investigate
the interaction between the hyperon and the nucleon
in the SU(3) Skyrme model.
The Atiyah-Manton ansatz extended to the SU(3) symmetry is adopted
as the two-baryon configuration.
The static potential is expanded in the modified SU(3) rotational matrices.
We obtain the interaction between the baryons
by integrating the static potential with the initial and final wave functions
over the Euler angles.
To obtain the attractive force in the central channel of the $\Lambda$-N interaction,
we take account of the $\Lambda$N-$\Sigma$N mixing through the intermediate state.

In Sec.\ II, we construct the two-baryon configuration
by the Atiyah-Manton ansatz.
In Sec.\ III, we express the potential
in the modified SU(3) rotational matrices
and obtain its matrix element between the baryons.
In Sec.\ IV, we consider the $\Lambda$N-$\Sigma$N mixing through the intermediate state
together with the finite $N_{\rm C}$ effects.
In Sec.\ V, we discuss our results.

\section{Two-Baryon Configuration}

Let us consider the non-linear field of the pseudo-scalar meson $ U $
within the flavor SU(3) symmetry.
The action of the SU(3) Skyrme model is given by
\begin{equation}
S=\int dt(L_2+L_4+L_{\rm SB})+N_{\rm C}\Gamma,
\label{eq:LYA}
\end{equation}
where
\begin{mathletters}
\begin{eqnarray}
L_2 & = & \int d^3x\frac{F_\pi^2}{16}
  {\rm tr}\left(\partial_\mu U^\dagger\partial^\mu U\right), \\
L_4 & = & \int d^3x\frac{1}{32e^2}
  {\rm tr}\left[U^\dagger(\partial_\mu U),U^\dagger(\partial_\nu U)\right]^2, \\
L_{\rm SB} & = & \int d^3x\Biggl\{
  \frac{F_\pi^2}{32}(m_\pi^2+m_\eta^2)
  {\rm tr}\left(U+U^\dagger-2\right) \nonumber \\
& & +\frac{\sqrt{3}F_\pi^2}{24}(m_\pi^2-m_{{\rm K}}^2)
  {\rm tr}\left(\lambda_8(U+U^\dagger)\right)\Biggr\},
\label{eq:LSB} \\
\Gamma & = & -\frac{i}{240\pi^2}\int_Q d^5x\epsilon^{ijklm}
  {\rm tr}\left(U^\dagger(\partial_i U)U^\dagger(\partial_j U)U^\dagger(\partial_k U)
 U^\dagger(\partial_l U)U^\dagger(\partial_m U)\right).
\label{eq:SWZ}
\end{eqnarray}
\end{mathletters}
The summation over the repeated indices is assumed
and $ \lambda_a $ denote the Gell-Mann matrices.
The symmetry breaking part of the lagrangian (\ref{eq:LSB}) reproduces
the mass terms expanded in the pseudo-scalar meson fields
with the Gell-Mann-Okubo relation
$ m_\pi^2+3m_\eta^2-4m_{\rm K}^2=0 $.
In the Wess-Zumino-Witten term (\ref{eq:SWZ}),
the integration is taken over the 5-dimensional disc $ Q $
the boundary of which is the usual spacetime.
The length and the meson mass are often measured in the unit $ 1/(eF_\pi)$,
and the energy in $ (F_\pi/e) $,
called the Skyrme units.

The hedgehog configuration is also stable in the SU(3) Skyrme model.
The quantization is done with respect to the collective coordinates
expressed as the rotation of the hedgehog configuration,
\begin{equation}
U(x,t)=A(t)U_{\rm H}(x)A^\dagger(t).
\label{eq:rothghg}
\end{equation}

It is difficult to construct the two-baryon configuration in the general form.
The product ansatz is used as the first approximation.
The product ansatz holds when the two skyrmions are separated in the long distance.

The Atiyah-Manton ansatz is another method
to construct the two-baryon configuration,
which has been used in the SU(2) Skyrme model.
It is obtained from the instanton,
a topological configuration of the gauge field
defined in the Euclidean spacetime,
\begin{equation}
U_{\rm AM}(x)=P\exp\left(-\int_{-\infty}^{\infty}dtA_4(x,t)\right).
\label{eq:AMansatz}
\end{equation}
The instanton configuration given by 'tHooft\cite{tH} is expressed as
\begin{eqnarray}
A_4 & = & \frac{i}{2}\tau\cdot\partial\log\rho, \label{eq:inst} \\
\rho & = & 1+\frac{\lambda_1^2}{(t-T_1)^2+(x-X_1)^2}
  +\frac{\lambda_2^2}{(t-T_2)^2+(x-X_2)^2},
\end{eqnarray}
where $(T_i,X_i)$ and $\lambda_i$ are the instanton coordinate and the spreading
of the $ i $-th instanton respectively.
Jackiw-Nohl-Rebbi (JNR) proposed the more general form of the instanton configuration\cite{JNR}.
The two-instanton superpotential is expressed as
\begin{equation}
\rho = \frac{\lambda_1^2}{(t-T_1)^2+(x-X_1)^2}
  +\frac{\lambda_2^2}{(t-T_2)^2+(x-X_2)^2}+\frac{\lambda_3^2}{(t-T_3)^2+(x-X_3)^2}.
\end{equation} 
From the skyrmion point of view,
it can describe the stable configuration with the torus shape.
However, it is difficult to apply the JNR form to the SU(3) Skyrme model
because of the complex relation
between the instanton parameters $ (T_i,X_i) $ and the skyrmion position.
Therefore, we concentrate our efforts on the 'tHooft form.

To apply the above method to our problem,
we extend the Atiyah-Manton ansatz to the SU(3) symmetry.
We can change Eq.\ (\ref{eq:inst}) into the form
\begin{equation}
A_4=-\frac{i}{2}\left(\tau_1\cdot\partial_{1}
  +\tau_2\cdot\partial_{2}\right)\log\rho,
\end{equation}
because the differentiation with respect to the spatial variables
can be separated into that with the instanton coordinates.
Now, we extend the gauge group from SU(2) to SU(3)
by replacing the SU(2) $ \tau $-matrices
with the generators of the SU(3) group
$ \tau_{1a}=A\lambda_a A^\dagger $
and $ \tau_{2a}=B\lambda_a B^\dagger $ different for each instanton coordinate,
\begin{eqnarray}
A_4 & = & -i\tau_1\cdot(x-X_1)\frac{\lambda_1^2 s_2^2}%
{s_1^2\{(s_1^2+\lambda_1^2)(s_2^2+\lambda_2^2)-\lambda_1^2\lambda_2^2\}} \nonumber \\
& & -i\tau_2\cdot(x-X_2)\frac{\lambda_2^2 s_1^2}%
{s_2^2\{(s_1^2+\lambda_1^2)(s_2^2+\lambda_2^2)-\lambda_1^2\lambda_2^2\}},
\label{eq:modAM}
\end{eqnarray}
where we have used the notations
\begin{mathletters}
\begin{eqnarray}
s_1^2 & = & (t-T_1)^2+(x-X_1)^2 \\
s_2^2 & = & (t-T_2)^2+(x-X_2)^2.
\end{eqnarray}
\end{mathletters}
The Atiyah-Manton configuration does not have
the exponential damping behavior of the massive meson
in the long distance.
We introduce the additional parameters for the exponential damping
by the substitution in Eq.\ (\ref{eq:modAM}),
\begin{eqnarray}
\lambda_i^2 & \rightarrow &
 \lambda_i^2(1+\mu_i|x-X_i|)e^{-\mu_i|x-X_i|}.
\label{eq:Lmod}
\end{eqnarray}
This substitution improves the long-distance behavior
of the Atiyah-Manton configuration for the massive case.
It corresponds to the hedgehog solution with the profile function
\begin{equation}
F(r)=\pi\left(1-\frac{r}{\sqrt{r^2+\lambda^2(1+\mu r)e^{-\mu r}}}\right).
\end{equation}
The long-distance behavior of the profile function leads to
\begin{equation}
F(r)\rightarrow\frac{\pi\lambda^2}{2r^2}(1+\mu r){e^{-\mu r}}.
\end{equation}
In spite of the above modifications,
the baryon number of the Atiyah-Manton configuration is still
conserved.
Indeed, the baryon number is confirmed to be two within 1\% discrepancy
by the numerical simulation.
We use the third parameter set in Ref.\ \cite{YA}
($F_{\pi}$=82.9MeV, $e$=4.87, $m_{\rm K}$=769MeV)
throughout this paper.
The subtraction of the vacuum-like energy does not matter
because we use the energy difference between the two configurations.
There is ambiguity in determining the separation of the two skyrmions
for the generated configuration.
We adopt the separation between the two baryons as
\begin{equation}
R=2\left[\int d^3xB_0\left(z^2-\frac{1}{2}(x^2+y^2)\right)\right]^{1/2},
\end{equation}
where $ B_0 $ is the baryon number density\cite{HOA}.
We perform the numerical simulation by taking the orientations
of the individual skyrmions as $ A=\sqrt{C}^\dagger $ and $ B=\sqrt{C} $
which ensures the symmetry under the exchange between the two skyrmions.
We determine the instanton parameters $ \lambda_{1,2}, \mu_{1,2} $.
From the symmetry under the exchange of the two skyrmions,
we require that they should be equal for the both skyrmions.
It turns out that $ \lambda=2.6, \mu=0.342 $
for the relative orientation $ C=1 $
and $ \lambda=2.2, \mu=1.369 $ for $ C=e^{-i(\pi/2)\lambda_4} $
by minimizing the static energy.
These parameters give the classical mass $ M=39.9 $ in the Skyrme unit
which is consistent with the exact value 39.849
estimated by the numerical simulation in Ref.\ \cite{YA}.
The static potentials $ V_1 $, $ V_2 $, $ V_3 $, and $ V_4 $
for the orientations $ C=1 $, $ \exp(-i(\pi/2)\lambda_2) $,
$ \exp(-i(\pi/2)\lambda_3) $, and $ \exp(-i(\pi/2)\lambda_4) $ respectively
are shown in FIG.\ 1.
%
%
\begin{center}
\setlength{\unitlength}{1in}
\begin{picture}(6.5,3.5)
\put(0,0.5){\makebox(6.5,3){\epsfysize=3in \epsfbox{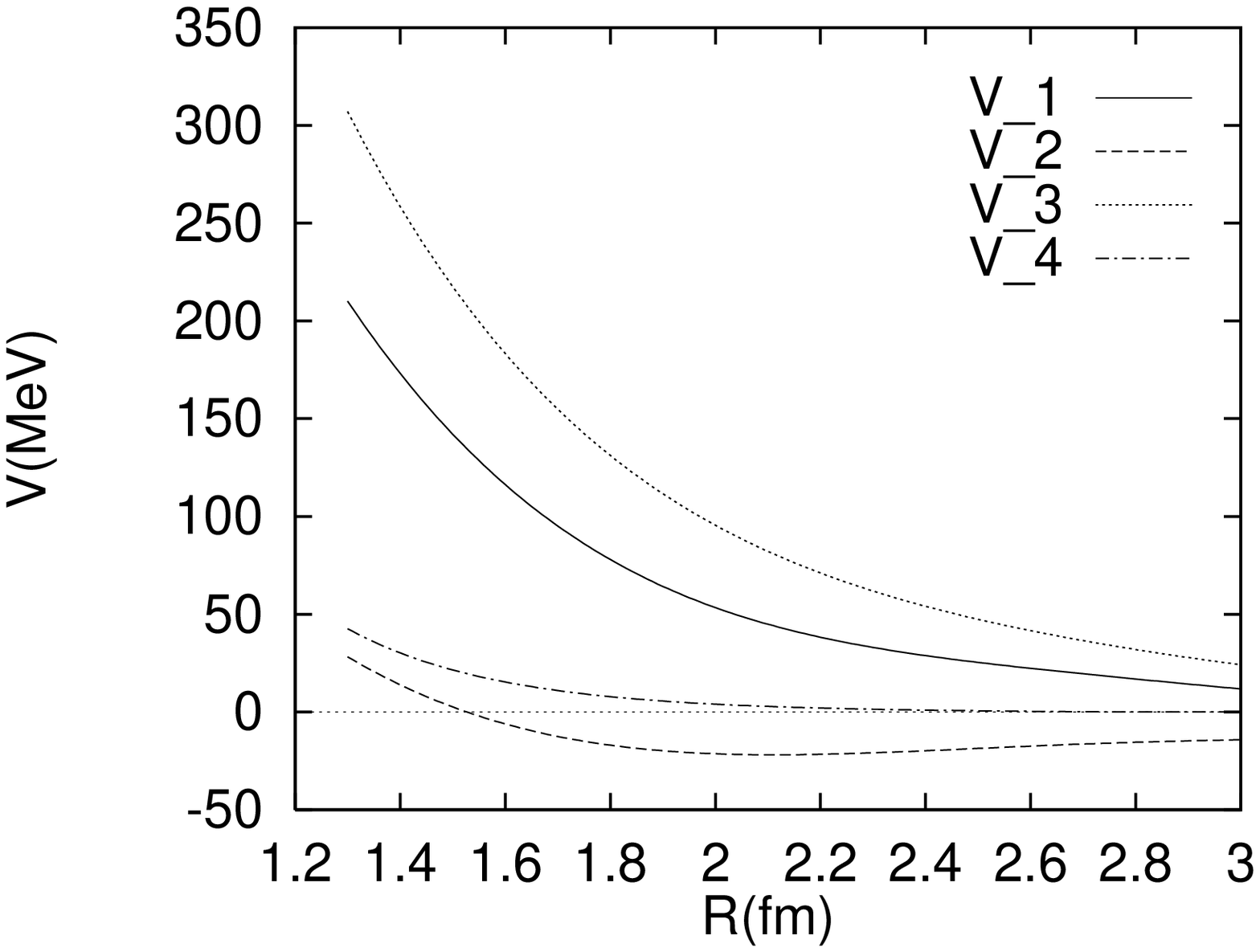}}}
\put(0,0){\makebox(6.5,0){\begin{minipage}{5in}FIG.\ 1 Static potential in MeV
as a function of the separation between the two baryons $ R $(fm)
$V_1$, $V_2$, $V_3$, $V_4$
for the relative orientation $C=1$, $e^{-(\pi/2)\lambda_2}$, $e^{-(\pi/2)\lambda_3}$,
$e^{-(\pi/2)\lambda_4}$.\end{minipage}}}
\end{picture}
\end{center}

\section{Hyperon-Nucleon Interaction}

The static potential is generally expanded
in the SU(3) rotation matrices\cite{SSG}.
From the symmetry of the solution,
the static potential is reduced to the form
\begin{equation}
V(R,C)=\sum_{\lambda,S_z}V^\lambda_{SS_z}(R)D^\lambda_{(0SS_z)(0SS_z)}(C),
\label{eq:Vexp}
\end{equation}
with the condition $ V^\lambda_{SS_z}=V^{\lambda^*}_{S,-S_z} $.

In the symmetry-breaking case,
it should be expanded in the modified rotation matrices
rather than the non-breaking ones
because the mixing of a certain representation
with its higher ones is not so small as to be neglected.
The SU(3) rotation is parameterized by the SU(3) Euler angles,
\begin{equation}
A = ue^{-i\nu\lambda_4}e^{-i(\rho/\sqrt{3})\lambda_8}u',
\label{eq:Eangle}
\end{equation}
where the matrices $ u $ and $ u' $ are expressed in the usual SU(2) Euler angles\cite{Holland}.
The wave function for the baryon belonging to the multiplet $ \lambda $
is given by Yabu and Ando as the modified SU(3) rotation matrix,
\begin{equation}
\Psi^\lambda_{(YII_z)(Y_{\rm R}SS_z)}(A)
=\sqrt{N_\lambda}\tilde{D}^\lambda_{(YII_z)(Y_{\rm R}S,-S_z)}(A),
\label{eq:YAwf}
\end{equation}
\begin{equation}
\tilde{D}^\lambda_{(YII_z)(Y_{\rm R}S,-S_z)}(A)
=D^I_{I_zM_{\rm L}}(u)
 f^\lambda_{(YIM_{\rm L})(Y_{\rm R}SM_{\rm R})}(\nu)
 e^{-i\rho Y_{\rm R}}D^S_{M_{\rm R},-S_z}(u'),
\end{equation}
where $ D^I_{I_zM_{\rm L}} $ and $ D^S_{M_{\rm R},-S_z} $
stand for the usual SU(2) rotation matrices
and $ N_\lambda $ is the multiplicity of the representation $\lambda$.
The properties of the symmetry breaking are contained
in the strange-mixing function $ f(\nu) $.
A subsidiary condition derived from the Wess-Zumino term
is imposed on the physical states,
\begin{equation}
Y_R=1.
\end{equation}
We can determine the coefficients $ V^\lambda_{SS_z} $
in the static potential (\ref{eq:Vexp})
by observing it for several relative orientations.
Once the static potential is given,
the interaction between the hyperon and the nucleon
is obtained by integrating the static potential between the initial and the final wave functions
over all orientations,
\begin{equation}
V^{\rm YN}(R)=\int dAdB\Psi_{Y'}^*(A)\Psi_{N'}^*(B)V(R,A^\dagger B)
 \Psi_Y(A)\Psi_N(B).
\end{equation}
where we adopt the direct product of the Yabu-Ando wave function
as the two-baryon state for the first approximation.
The interaction is obtained in the form
\begin{eqnarray}
V^{\rm NN} & = & V^{\rm NN}_{\rm C}+V^{\rm NN}_{\tau}(\tau_1\cdot\tau_2)
+V^{\rm NN}_{\sigma}(\sigma_1\cdot\sigma_2)
+V^{\rm NN}_{\rm T}S_{12} \nonumber \\
& & +V^{\rm NN}_{\sigma\tau}(\sigma_1\cdot\sigma_2)(\tau_1\cdot\tau_2)
+V^{\rm NN}_{\rm T\tau}S_{12}(\tau_1\cdot\tau_2),
\end{eqnarray}
for the NN-interaction,
\begin{equation}
V^{\Lambda \rm N}=V^{\Lambda \rm N}_{\rm C}
+V^{\Lambda \rm N}_{\sigma}(\sigma_1\cdot\sigma_2)
+V^{\Lambda \rm N}_{\rm T}S_{12},
\end{equation}
for the $\Lambda$N-interaction,
and
\begin{eqnarray}
V^{\Sigma \rm N} & = & V^{\Sigma \rm N}_{\rm C}
+V^{\Sigma \rm N}_{\tau}(T_1\cdot\tau_2)
+V^{\Sigma \rm N}_{\sigma}(\sigma_1\cdot\sigma_2) \nonumber \\
& & +V^{\Sigma \rm N}_{\rm T}S_{12}
+V^{\Sigma \rm N}_{\sigma\tau}(\sigma_1\cdot\sigma_2)(T_1\cdot\tau_2)
+V^{\Sigma \rm N}_{\rm T\tau}S_{12}(T_1\cdot\tau_2),
\end{eqnarray}
for the $\Sigma$N-interaction.
Since the baryons stay in the $z$-axis,
the tensor operator is defined
by $ S_{12}=3\sigma_{1z}\sigma_{2z}-\sigma_1\cdot\sigma_2 $.
The potentials $V_1, V_2, V_3, V_4 $ for the relative orientations
determine the coefficients $V^1_{00}$, $V^8_{00}$, $V^8_{10}$, $V^8_{11}$
in the static potential (\ref{eq:Vexp})
expanded up to the octet representation.
The graphs of the interaction $ V^{\rm NN}_{\rm C} $, $ V^{\rm NN}_{\sigma\tau} $,
$ V^{\rm NN}_{\rm T\tau} $, $ V^{\Lambda \rm N}_{\rm C} $, $ V^{\Sigma \rm N}_{\rm C} $,
$ V^{\Sigma \rm N}_{\sigma\tau} $, and $ V^{\Sigma \rm N}_{\rm T\tau} $
are shown in FIGs.\ 2-8.
\begin{center}
\setlength{\unitlength}{1in}
\begin{picture}(6.5,6)
%
%
\put(0,4){\makebox(6.5,2){\epsfysize=2in \epsfbox{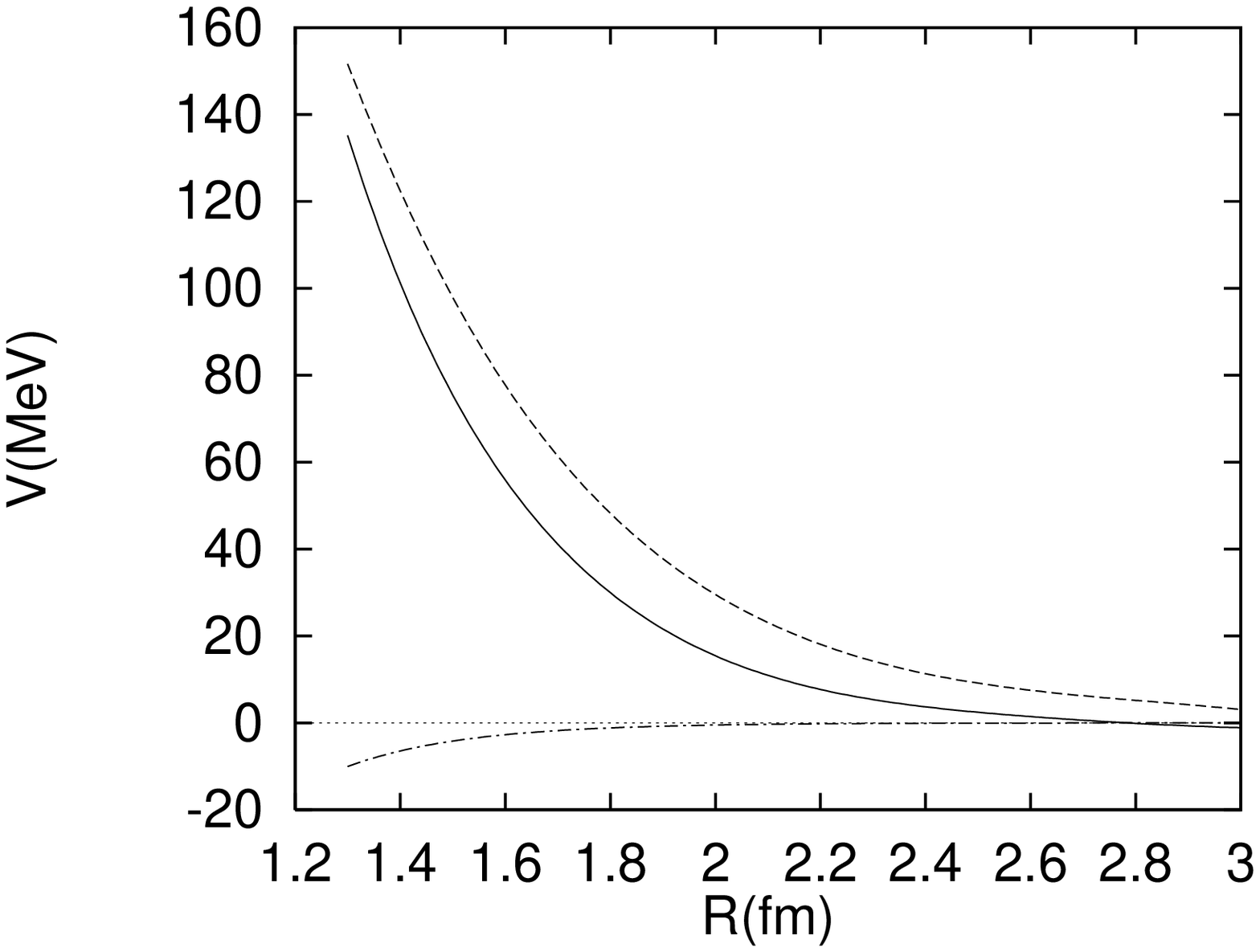}}}
\put(0,3){\makebox(6.5,0.5){\begin{minipage}{4in}
FIG.\ 2 Central part of the NN-potential $ V_{\rm C} $ in MeV
as a function of the separation $R$(fm),
dashed curve denotes the product ansatz
and dot-dashed one the one-boson exchange model.
\end{minipage}}}
%
%
\put(0,0.5){\makebox(3.25,2){\epsfysize=2in \epsfbox{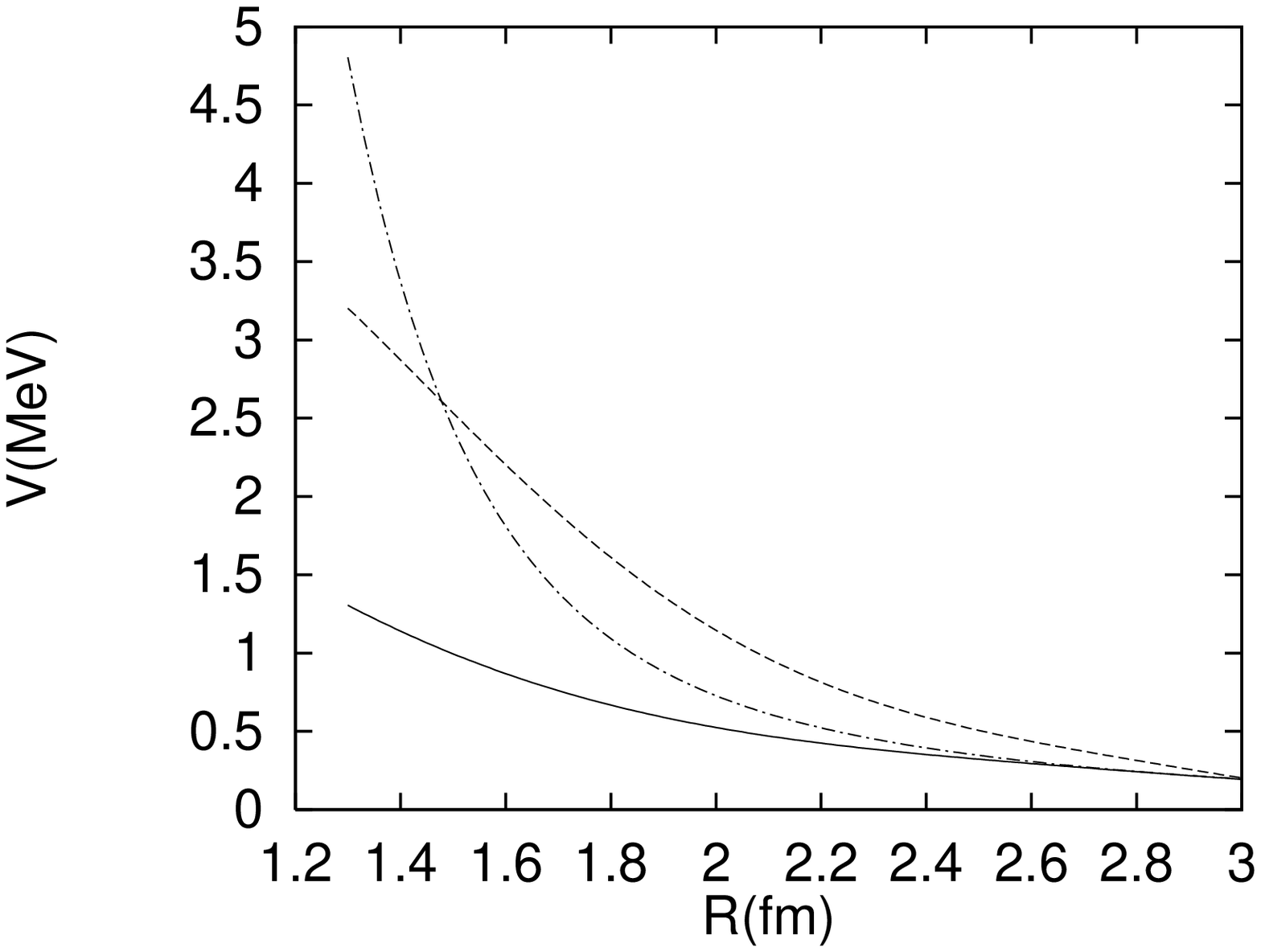}}}
\put(0,0){\makebox(3.25,0.5){\begin{minipage}{2.75in}
FIG.\ 3 Spin-isospin part of the NN-potential $ V_{\sigma\tau} $.
\end{minipage}}}
%
%
\put(3.25,0.5){\makebox(3.25,2){\epsfysize=2in \epsfbox{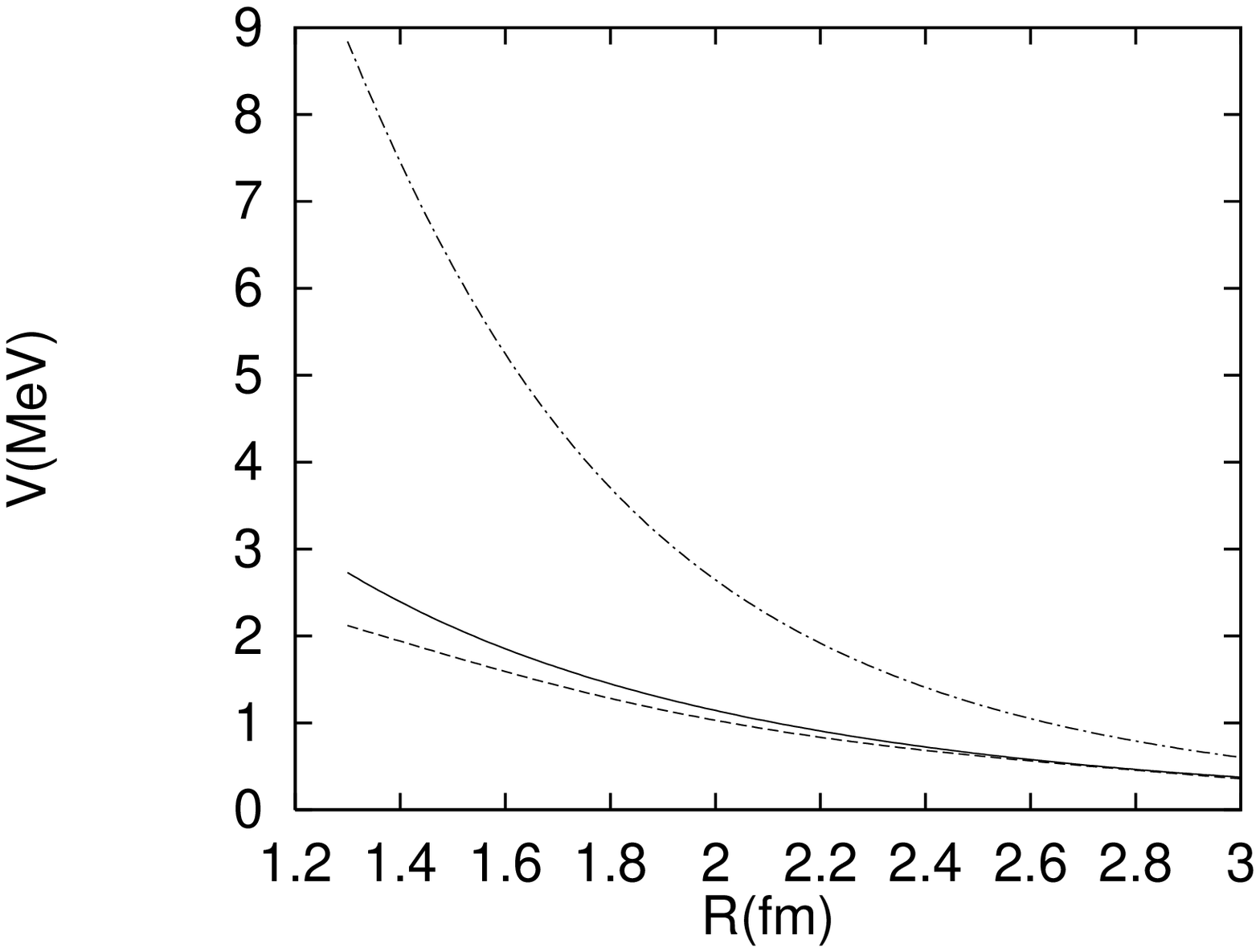}}}
\put(3.25,0){\makebox(3.25,0.5){\begin{minipage}{2.75in}
FIG.\ 4 Tensor-isospin part of the NN-potential $ V_{\rm T\tau} $.
\end{minipage}}}
\end{picture}
\end{center}
\begin{center}
\setlength{\unitlength}{1in}
\begin{picture}(6.5,5)
%
%
\put(0,3){\makebox(3.25,2){\epsfysize=2in \epsfbox{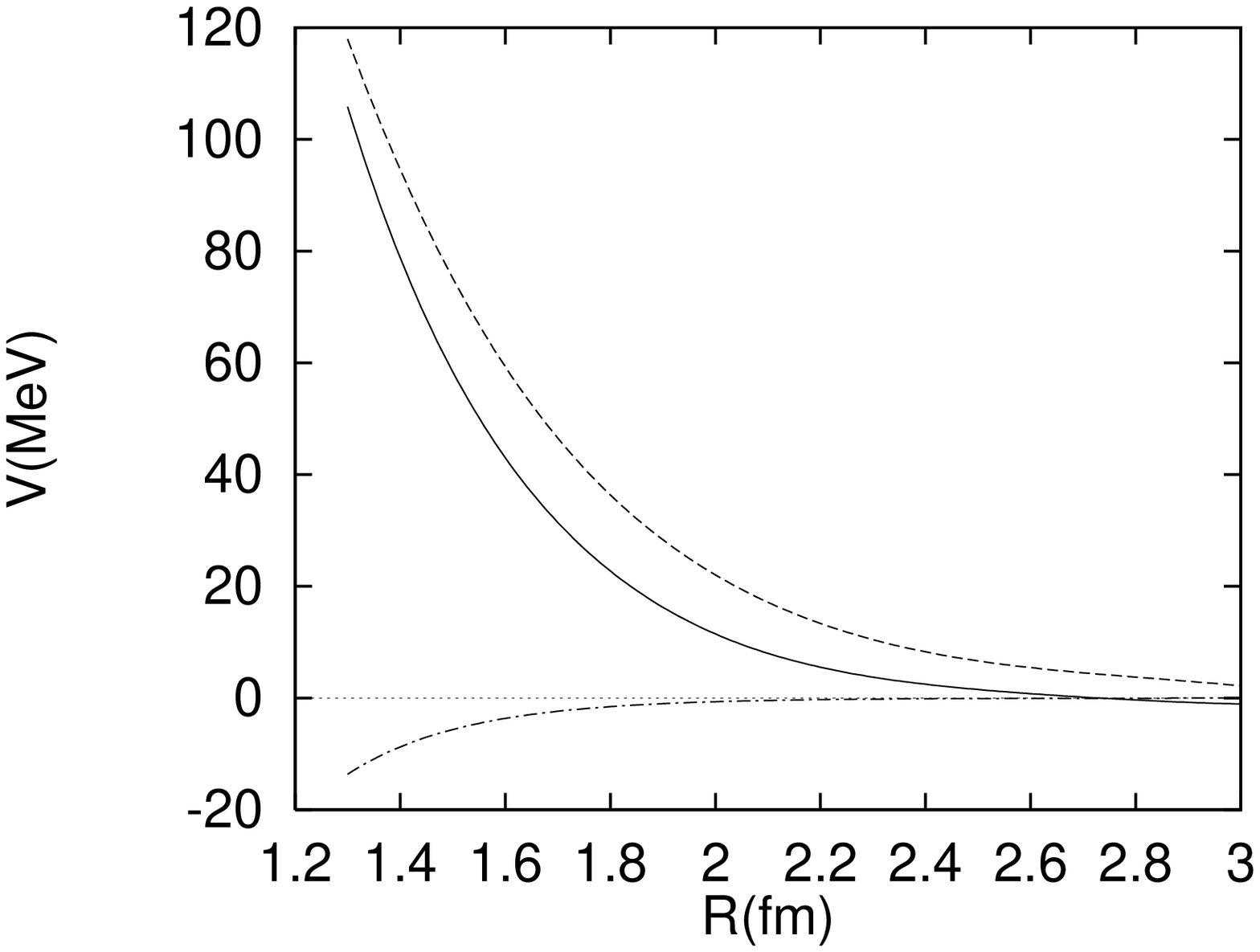}}}
\put(0,2.5){\makebox(3.25,0.5){\begin{minipage}{2.75in}
FIG.\ 5 Central part of the $\Lambda $N-potential $ V_{\rm C} $.
\end{minipage}}}
%
%
\put(3.25,3){\makebox(3.25,2){\epsfysize=2in \epsfbox{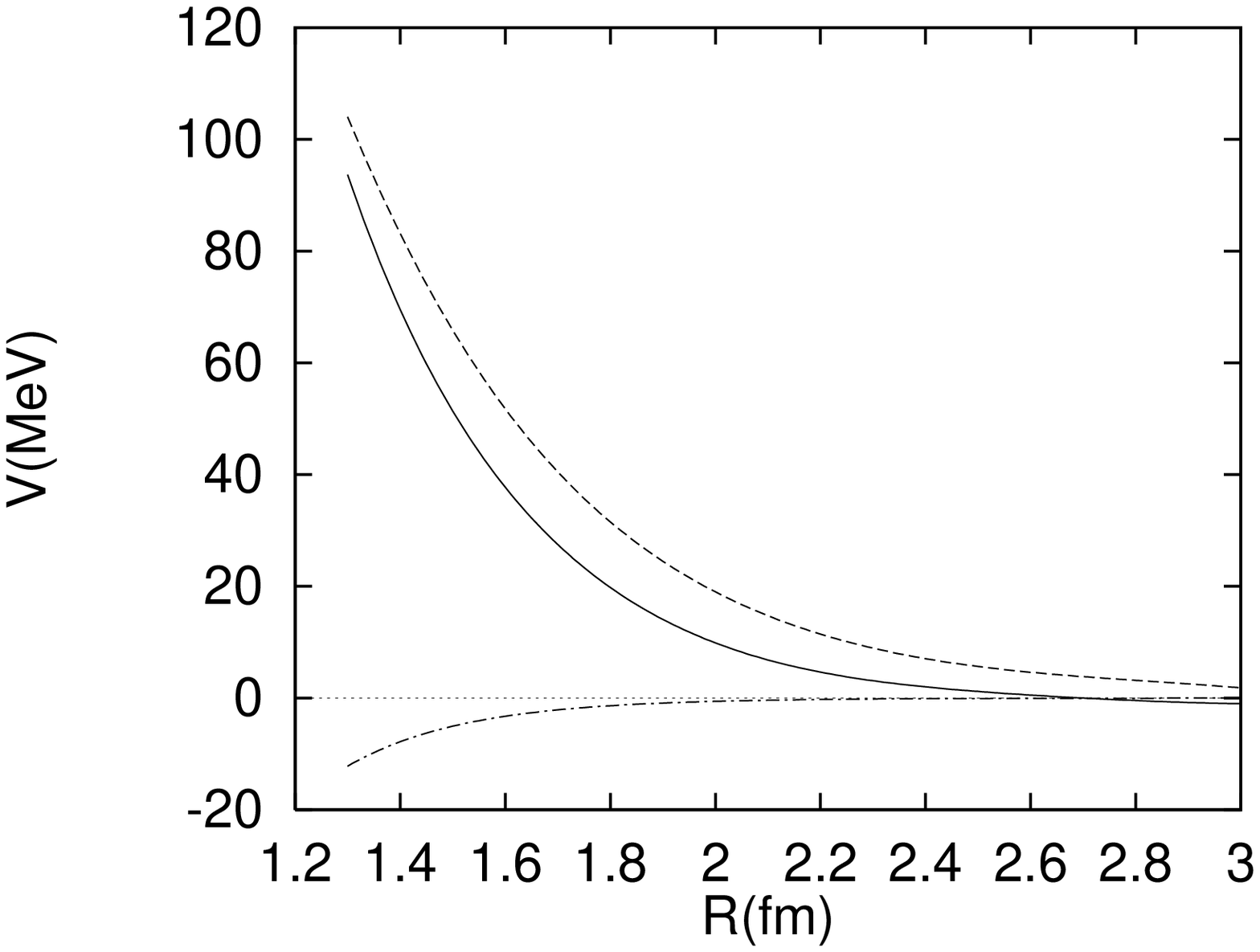}}}
\put(3.25,2.5){\makebox(3.25,0.5){\begin{minipage}{2.75in}
FIG.\ 6 Central part of the $\Sigma $N-potential $ V_{\rm C} $.
\end{minipage}}}
%
%
\put(0,0.5){\makebox(3.25,2){\epsfysize=2in \epsfbox{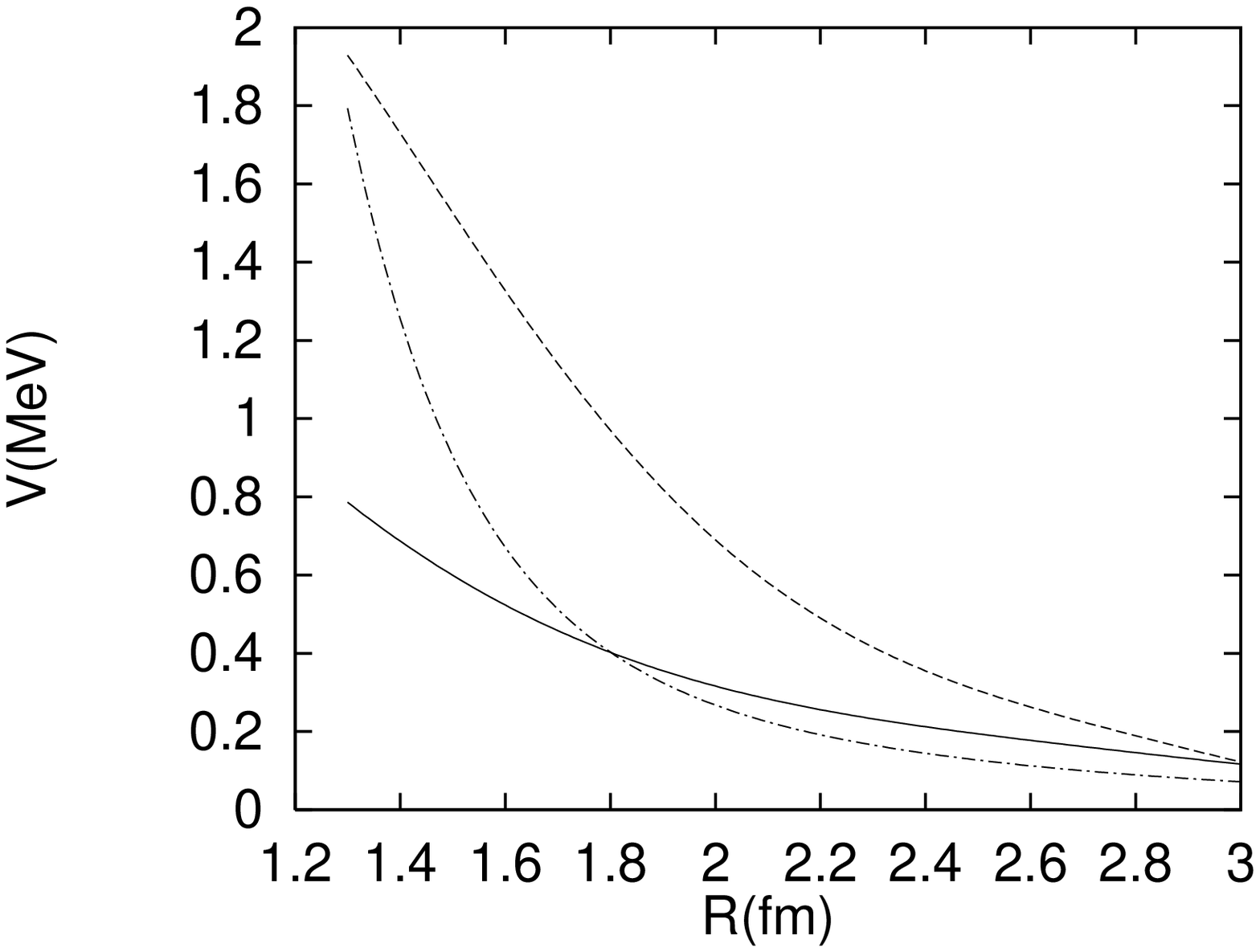}}}
\put(0,0){\makebox(3.25,0.5){\begin{minipage}{2.75in}
FIG.\ 7 Spin-isospin part of $\Sigma $N-potential $ V_{\sigma\tau} $.
\end{minipage}}}
%
%
\put(3.25,0.5){\makebox(3.25,2){\epsfysize=2in \epsfbox{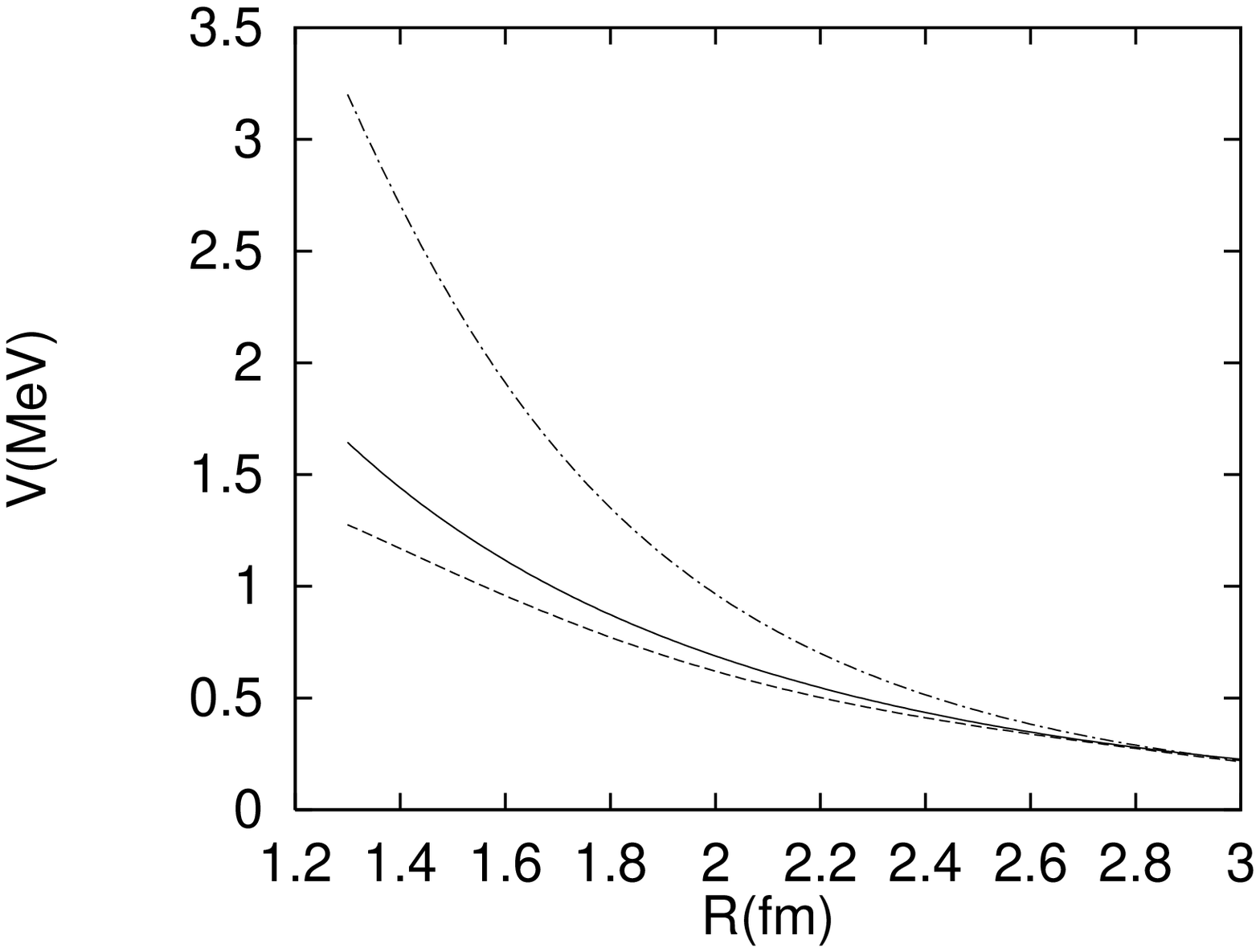}}}
\put(3.25,0){\makebox(3.25,0.5){\begin{minipage}{2.75in}
FIG.\ 8 Tensor-isospin part of the $\Sigma $N-potential $ V_{\rm T\tau} $.
\end{minipage}}}
\end{picture}
\end{center}
The central potential $ V_{\rm C} $ of the NN, $\Lambda$N, $\Sigma$N systems
is still repulsive,
while the Atiyah-Manton ansatz tends to show the less repulsive force
than the product ansatz \cite{KE,SSG}.
For the spin-isospin part $ V_{\sigma\tau} $,
the results show a good agreement
with the Nijmegen potential (model D),
the one-boson-exchange potential
developed by the Nijmegen group\cite{NRd}
at the range $ R>1.6{\rm fm} $.
The behavior of the tensor-isospin part $ V_{\rm T\tau} $
is consistent with the Nijmegen model.

\section{$\Lambda$N-$\Sigma$N Mixing}

It is well known that the naive estimation of the interaction is insufficient
to give the central attraction at the intermediate range.
In the SU(2) case, the $\Delta$-N mixing is taken into account.
The direct product of the wave functions as the two-baryon eigenstate
becomes worse when the separation between the skyrmions decreases.
The candidate for the SU(3) symmetry
is the $\Lambda$-$\Sigma$ mixing.
We consider the $\Lambda$N-$\Sigma$N mixing in the intermediate state
together with the finite $N_{\rm C}$ effects.
It is shown
by the fact that the off-diagonal element of the potential survives,
\begin{eqnarray}
V^{\Lambda{\rm N}-\Sigma{\rm N}} & = &
\left(V^{\Lambda{\rm N}-\Sigma{\rm N}}_{\tau}
+V^{\Lambda{\rm N}-\Sigma{\rm N}}_{\sigma\tau}(\sigma_1\cdot\sigma_2)
+V^{\Lambda{\rm N}-\Sigma{\rm N}}_{\rm T\tau}S_{12}\right)O_{\rm T},
\end{eqnarray}
where
\begin{equation}
O_{\rm T}=-\sqrt{2/3}T_{2+}\delta_{I_{1z}',1}+2/\sqrt{3}T_{2z}\delta_{I_{1z}',0}
+\sqrt{2/3}T_{2-}\delta_{I_{1z}',-1}.
\label{eq:OTdef}
\end{equation}
It is found that the isospin is conserved under the baryon-baryon interaction.
The total spin is not an invariant of the hyperon-nucleon system
whereas the projection of the spin in the $z$-direction is still conserved.
The potentials with respect to the total spin and isospin states
are written as
\begin{mathletters}
\begin{eqnarray}
V_{\Lambda\Lambda}
& = & V^{\rm \Lambda N}_{\rm C}-3V^{\rm \Lambda N}_{\sigma} \\
V_{\Sigma\Sigma}
& = & (V^{\rm \Sigma N}_{\rm C}-2V^{\rm \Sigma N}_{\tau})
-3(V^{\rm \Sigma N}_{\sigma}-2V^{\rm \Sigma N}_{\sigma\tau}) \\
V_{\Lambda\Sigma}
& = & \frac{5}{3}(V^{\rm \Lambda N-\Sigma N}_{\tau}
-3V^{\rm \Lambda N-\Sigma N}_{\sigma\tau}),
\end{eqnarray}
\end{mathletters}
for the $ I=\frac{1}{2}, S=0 $ channel,
\begin{mathletters}
\begin{eqnarray}
V_{\Lambda\Lambda}
& = & V^{\rm \Lambda N}_{\rm C}+V^{\rm \Lambda N}_{\sigma}-4V^{\rm \Lambda N}_{\rm T} \\
V_{\Sigma\Sigma}
& = & (V^{\rm \Sigma N}_{\rm C}-2V^{\rm \Sigma N}_{\tau})
+(V^{\rm \Sigma N}_{\sigma}-2V^{\rm \Sigma N}_{\sigma\tau})
-4(V^{\rm \Sigma N}_{\rm T}-2V^{\rm \Sigma N}_{\rm T\tau}) \\
V_{\Lambda\Sigma}
& = & \frac{5}{3}(V^{\rm \Lambda N-\Sigma N}_{\tau}
+V^{\rm \Lambda N-\Sigma N}_{\sigma\tau})
-4V^{\rm \Lambda N-\Sigma N}_{\rm T\tau},
\end{eqnarray}
\end{mathletters}
for the $ I=\frac{1}{2}, S=1, S_z=0 $ channel, and
\begin{mathletters}
\begin{eqnarray}
V_{\Lambda\Lambda}
& = & V^{\rm \Lambda N}_{\rm C}+V^{\rm \Lambda N}_{\sigma}+2V^{\rm \Lambda N}_{\rm T} \\
V_{\Sigma\Sigma}
& = & (V^{\rm \Sigma N}_{\rm C}-2V^{\rm \Sigma N}_{\tau})
+(V^{\rm \Sigma N}_{\sigma}-2V^{\rm \Sigma N}_{\sigma\tau})
+2(V^{\rm \Sigma N}_{\rm T}-2V^{\rm \Sigma N}_{\rm T\tau}) \\
V_{\Lambda\Sigma}
& = & \frac{5}{3}(V^{\rm \Lambda N-\Sigma N}_{\tau}
+V^{\rm \Lambda N-\Sigma N}_{\sigma\tau})
+2(V^{\rm \Lambda N-\Sigma N}_{\rm T\tau}),
\end{eqnarray}
\end{mathletters}
for the $ I=\frac{1}{2}, S=1, S_z=\pm 1 $ channel.
The non-zero off-diagonal matrix element $ V^{\rm \Lambda N-\Sigma N} $
shows that the direct product of the single-baryon wave functions
is not a good eigenstate for the two-baryon system.
One can obtain the better two-baryon state
by diagonalizing the matrix
$\left(\begin{tabular}{cc}
$V_{\Lambda\Lambda}$ & $V_{\Lambda\Sigma}$ \\
$V_{\Lambda\Sigma}$ & $V_{\Sigma\Sigma}$ \\
\end{tabular}\right)$ for each channel.
After the diagonalization,
the lowest eigenvalue is adopted for the $\Lambda$N central potential.
The finite $N_{\rm C}$ effects should be taken into consideration
together with the $\Lambda$-$\Sigma$ mixing
because the Skyrme model is recognized as an effective theory
in the large $N_{\rm C}$ limit.
The spin-isospin matrix elements are enhanced in the $N_{\rm C}=3$ case
compared with those in the large $N_{\rm C}$ limit
by the factors 20/9 for $V^{\rm \Sigma N}_{\sigma\tau}$, $V^{\rm \Sigma N}_{\rm T\tau}$
and $(20\sqrt{3})/9$ for $V^{\rm \Lambda N-\Sigma N}_{\sigma\tau}$,
$V^{\rm \Lambda N-\Sigma N}_{\rm T\tau}$,
from the analysis of the quark hedgehog model as in Ref.\ \cite{JJP}.
The graph of the $\Lambda$-N interaction
in the central channel $ V^{\Lambda \rm N}_{\rm C} $
with the $\Lambda$N-$\Sigma$N mixing is shown
in FIG.\ 9.
\begin{center}
\setlength{\unitlength}{1in}
\begin{picture}(6.5,4)
%
%
\put(0,1){\makebox(6.5,3){\epsfysize=3in \epsfbox{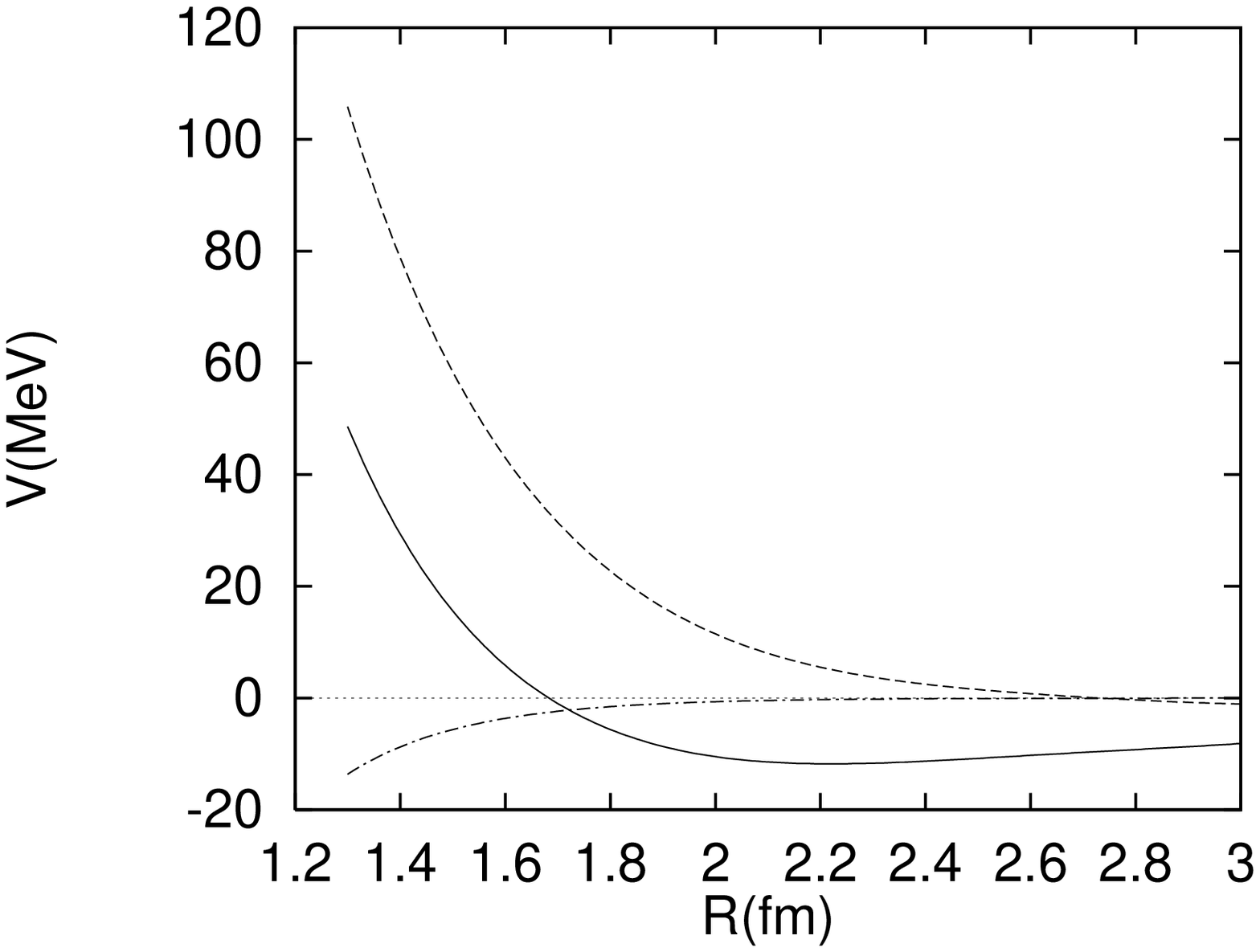}}}
\put(0,0){\makebox(6.5,0.5){\begin{minipage}{5in}
FIG.\ 9 Central part of the $\Lambda $N-potential $ V_{\rm C} $
with $\Lambda$N-$\Sigma$N mixing,
solid curve stands for that with the $\Lambda$-$\Sigma$ mixing,
dashed one for the naive estimation,
and dot-dashed one for the one-boson-exchange model.
\end{minipage}}}
\end{picture}
\end{center}
The attractive force at the intermediate range appears
by taking account of the $\Lambda$N-$\Sigma$N mixing through the intermediate state
together with the finite $N_{\rm C}$ effects.

\section{Discussion}

We discuss the results in this section.
The static potential is obtained from the Atiyah-Manton ansatz
extended to the SU(3) symmetry.
The Atiyah-Manton ansatz gives a lower energy
than the product ansatz at the intermediate range.
Since the symmetry breaking is not small,
the static potential is expanded
in the modified SU(3) rotation matrices
up to the octet representation.
We have obtained the baryon-baryon interaction
by integrating the static potential between the two-baryon states
over the Euler angles.
In the naive estimation,
we have not obtained the intermediate attraction of the central force
although the result from the Atiyah-Manton ansatz is less repulsive
than the product ansatz.
To improve the estimation of the central potential,
we have to take account of the several effects as in the SU(2) case.
One of such effects is introduced by considering the mixing with the higher excitations,
the $\Delta$-N mixing in the SU(2) case,
through the intermediate state.
The candidate within the SU(3) symmetry
is the mixing between $ \Sigma $ and $ \Lambda $\cite{Gal}.
The effects from the mixing of these particles are expected to play a significant role
in the hyperon-nucleon interaction
because the mass difference between $\Lambda$ and $\Sigma$,
$ m_\Sigma-m_\Lambda\simeq 80{\rm MeV} $,
is smaller than that between N and $ \Delta $,
$ m_\Delta-m_{\rm N}\simeq 300{\rm MeV} $.

The central potential between $\Lambda$ and N with the $\Lambda$N-$\Sigma$N mixing
shows the attraction at the intermediate range.
This result is consistent with the one-boson-exchange model.
The direct product of the two single-baryon states is not a good eigenstate
when the two skyrmions close together.
It is suggested
by the non-vanishing off-diagonal matrix element between the $\Lambda$-N and $\Sigma$-N states.
By diagonalizing this matrix for each channel,
one can obtain the better eigenstate of the two-baryon system.
The lowest eigenvalue is adpoted for the $\Lambda$N potential.
This procedure amounts
to taking account of the $\Lambda$N-$\Sigma$N mixing through the intermediate state.
The finite $N_{\rm C}$ effects is considered together with the $\Lambda$-$\Sigma$ mixing.
The finite $N_{\rm C}$ correction is estimated from the analysis of quark hedgehog model.

The spin-isospin part $ V_{\sigma\tau} $
and the tensor-isospin part $ V_{\rm T\tau} $
shows a consistent behavior with the one-boson-exchange model
at the range $ R>1.6{\rm fm} $.
This implies that the long-range force which is dominated by the $\pi$-exchange
reproduces the one-boson-exchange potential well.

In the present paper,
we have observed
that the naive estimation of the interaction between the hyperon and the nucleon
does not show the attractive central force at the intermediate range.
The Atiyah-Manton ansatz is adopted to improve the medium-range behavior
of the skyrmion configuration rather than the product ansatz.
The $\Lambda$N-$\Sigma$N mixing in the intermediate state is taken into account.
The finite $N_{\rm C}$ effects are included
from the quark hedgehog model.
After these treatments are taken,
the hyperon-nucleon interaction
shows the central attraction which is consistent with the one boson exchange model.
Therefore, we conclude that the configuration with a certain accuracy,
the $\Lambda$N-$\Sigma$N mixing,
and the finite $N_{\rm C}$ effects
are required for the attractive force in the $\Lambda$-N interaction.

Finally, we discuss the validity of the Atiyah-Manton configuration
based on the 'tHooft instanton.
In this paper, we have adopted the 'tHooft form
as a starting point of the SU(3) skyrmion configuration.
On the other hand, the Jackiw-Nohl-Rebbi form gives the more general configuration.
It can describe the stable configuration with the torus shape.
Indeed, it is significant to reproduce such a configuration
in the SU(3) model as well.
In this field,
the stable point in the manifold of the Atiyah-Manton configuration
is investigated.
By quantizing the fluctuation around it,
one constructs the quantum state which has the same quantum number
as the deuteron\cite{LMSW}.
However, the JNR form is difficult to handle the modification of the long-distance behavior
caused by the mass of the pseudo-scalar meson.
At this point, it is convenient to take the 'tHooft form
owing to the transparent relation
between the instanton coordinate and that of the skyrmion.
Furthermore, our configuration based on the 'tHooft form
is still valid in the region
where the individual skyrmions are identified.

\section*{Acknowledgment}

The author thanks Professor K.\ Ohta for his suggestion to investigate this subject.
\end{document}